\documentclass[twocolumn, 10pt, floatfix, superscriptaddress, booktabs]{revtex4}

\usepackage{amsmath}
\usepackage{amssymb}
\usepackage{amsfonts}   
\usepackage{graphicx}   
\usepackage{verbatim}   
\usepackage{color}      
\usepackage{subfigure}
\usepackage{pgfplotstable}
\usepackage{wrapfig}
\usepackage{hhline}

\usepackage{tikz}
\usetikzlibrary{positioning}
\definecolor{myRed}{RGB}{229,25,50}
\definecolor{myBlue}{RGB}{25,178,255}
\definecolor{myGreen}{RGB}{50,255,0}

\begin{document}

\title{Correlated Avalanche Burst Invasion Percolation: Multifractal origins of self organized critiality}
\date{\today}

\author{Ronaldo Ortez}
\email{raortez@ucdavis.edu}
\affiliation{Department of Physics, One Shields Ave., University of California, Davis, CA 95616, United States}
\author{John B. Rundle}
\email{rundle@ucdavis.edu}
\affiliation{Department of Physics, One Shields Ave., University of California, Davis, CA 95616, United States}
\affiliation{Department of Geology, One Shields Ave., University of California, Davis, CA 95616, United States}
\affiliation{Santa Fe Institute, Santa Fe, NM 87501, United States}

\begin{abstract}
	We extend our previous model, avalanche-burst invasion percolation (AIP) by introducing long-range correlations between sites described by fractional Brownian statistics. In our previous models with independent, random site strengths, we reproduced a unique set of power-laws consistent with some of the b-values observed during induced seismicity. We expand upon this model to produce a family of critical exponents which could be characterized by the local long-range correlations inherent to host sediment. Further, in previous correlated invasion percolation studies, fractal behavior was found in only a subset of the range of Hurst exponent, $H$. We find fractal behavior persists for the entire range of Hurst exponent. Additionally, we show how multiple cluster scaling power laws results from changing the generalized Hurst parameter controlling long-range site correlations, and gives rise to a multifractal system. This emergent multifractal behavior plays a central role in allowing us to extend our model to better account for variations in the observed Gutenber-Richter b-values of induced seismicity. 
\end{abstract}

\maketitle

\section{Introduction}
One of the most interesting insights from random percolation (RP) is the emergence of long range correlations from the inherently random process of independently occupying sites with probability $p$ on a lattice. Much of percolation's value comes from providing an extremely simple framework from which many puzzling features both can arise and can be understood with emergent scale invariant connectivity chief among these. In addition to emergent long-range correlations near the critical point, researchers became curious about the effects of implicit long-range lattice structure correlations on critical behavior\cite{weinrib1984long,prakash1992structural,schrenk2013percolation,sahimi1996scaling,makse1996method}. This question is of interest not only from a formal perspective, but also, because long-range correlations (LRC) are described by fractal relations, which now account for a large number and variety of natural systems \cite{mandelbrot1982fractal}.

In our previous paper \cite{Ortez2022critical}, we characterized the pseudo-critical behavior of our avalanche-burst invasion percolation (AIP) model, which produced a critical distribution of bursts, $n_s(T)$, as a function of strength threshold, $T$. AIP's stochastic growth mechanism reproduces a distribution of invaded sites consistent with percolation's emergent long range order so as to produce a unique burst distribution characterized by exponents, $\tau=1.594 \pm.009$, $\sigma=0.41\pm.01$. These exponents are near but distinct from mean-field cluster scaling,$\tau_{MF}=1.5, \sigma=0.5$, and coupled with the correlation scaling of sites within bursts, $\xi_b \sim \epsilon_T ^{-\nu}$ ($\nu=1.3$) serves to define a distinct universality class of critical behavior, distinct even from RP. 

Only a few studies have been done on LRC on IP \cite{knackstedt2000invasion,Vidales_1996}, and these studies only looked at the static network type scaling ($D_f,D_{min},D_b$) properties which in this case do little to provide insight into how the critical properties change. This is of course largely because the critical description of IP has been poorly understood, and had not been placed within the appropriate framework to assign it various critical properties. This was done with our AIP model, and now positions us to address the topic of LRC's impact on AIP's pseudo-critical behavior. 

In addition to these theoretical considerations, there are the more phenomenological ones. In particular, our AIP model is a characteristic self organized critical (SOC) type system with slowly driven non-equilibrium dynamics that result in effective power-law behavior. Robust definitions of SOC remain elusive as is their connection to critical behavior \cite{turcotte2001self}, and this effort aims at establishing the connection between traditional critical processes (characterized by a single correlation length) and SOC systems (likely containing multiple correlation lengths in the system). Special interest is in the seismic applications of SOC and our model specifically aims at reproducing the Gutenber-Richter scaling consistent with induced seismicity \cite{ortez2021universality}. Beyond the possibility that instabilities in stress field can be triggered by small fluctuations to self organizing behavior \cite{grasso1998testing}, we model the infiltration of invading fluid into a defending substrate as a slowly driven invasion percolation process following a principle of least resistance through a lattice of sites with random and isotropic resistance. The invasion path will naturally select the subset of sites where we can observe long range correlations between the invaded sites. The additional burst mechanism allows us to identify the conditions which yield scale invariant bursts, and thus, allows us to speculate on the conditions that must exist to produce the observed scale invariant seismic distributions. 

This application is made more accurate because studies of porous media find correlations between pore size in various sedimentary substrates. These studies indicate porous media “sites” are not independent and random, but rather, exhibit long-range correlations. In particular, fractional-brownian statistics seem to well describe the porosity logs within many heterogeneous rock formations at large scales\cite{knackstedt1998simulation}. Similar findings for the permeability distribution have been found for oil reservoirs and aquifers \cite{leary2002power}. Therefore, in this paper, we show how the characteristics of our AIP algorithms change in the presence of implicitly correlated lattice sites rather than a lattice of independent random sites. 

\begin{figure*}[t] 
	\centering
	\includegraphics[width=1.0\textwidth]{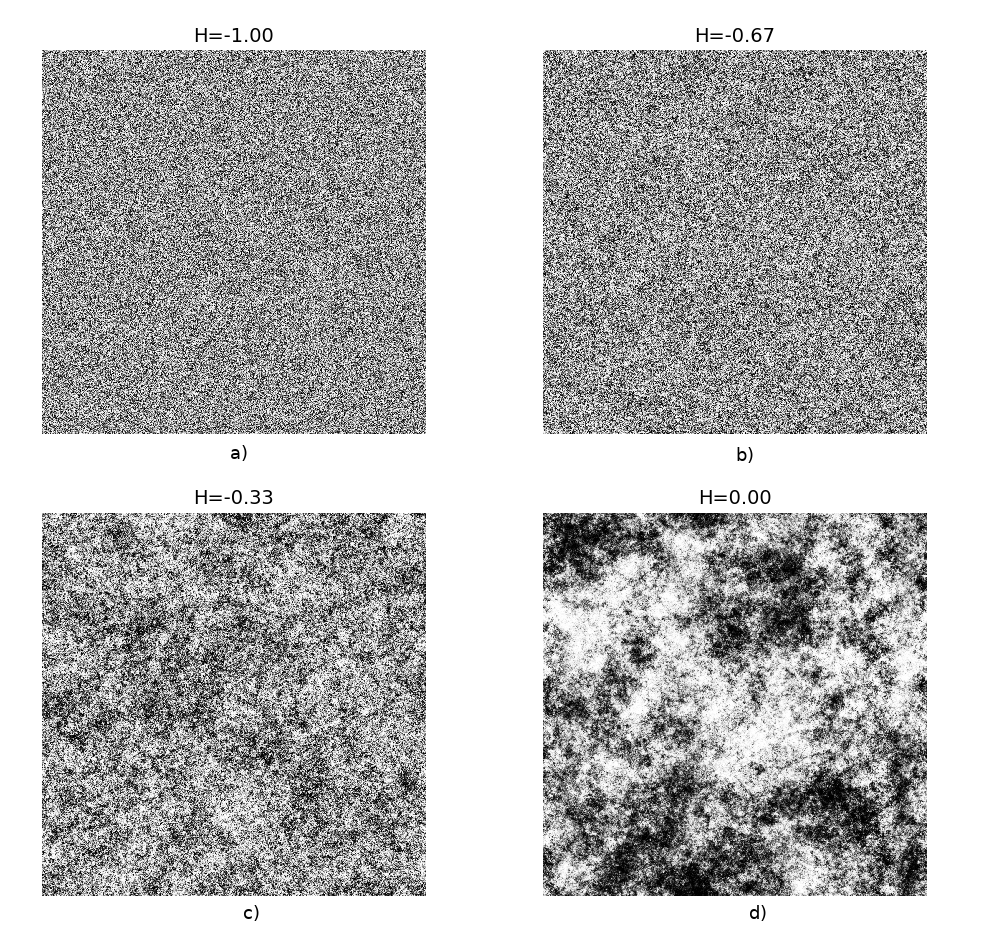}
	\caption{\footnotesize Sampling of lattices with increasing correlation. We show how the lattice sites become increasingly correlated as the generalized Hurst exponent increases from -1 to 0. a) $H=-1.0$ corresponds to the random case. b) $H=-0.67$ corresponds to antipersistent correlations c) $H=-0.33$ corresponds to persistent correlations d) $H=0.0$ corresponds to increasingly large correlations where clustering of similar strengths is clearly observable}
	\label{fig:corrLattEx}
\end{figure*}

\section{Long Range Correlations}
\label{section:LRC}
It is common to parameterize long range correlations using the Hurst exponent, $H$, where the (auto)correlation function, $C(r)$ defined as $C(r)=\left\langle u(r^\prime)u(r+r^\prime)\right\rangle$ has the following behavior:
\begin{equation}
C(r) \propto r^{2H}
\label{eq:hurstCorr}
\end{equation}
where $H$ is taken to be in the range [0,1].
Harris provided a powerful framework for anticipating the effect that changes in lattice structure could have on subsequent behavior. Weinrib \cite{weinrib1984long} extended Harris' formulation specifically to the percolation problem. We can largely adopt much of the existing framework, where we recognize that AIP's critical behavior is described by a critical control parameter, burst threshold $T$, rather than a critical occupation probability, $p_c$. This means that fluctuations in occupation probabilities correspond to fluctuations in bursts described by $T$.

We give the derivation in Appendix \ref{appedix:LRC} which traditionally considers well behaved correlations of the type $C(r)\sim r^{-a}$, and the condition on $a$ such that the system preserves the existence of a uniform transition. This is given by,
\begin{equation}
a\nu - 2 > 0
\label{eq:HarrisCond}
\end{equation} 
Thus, we can expect changes to the critical behavior if $a<2/\nu$. For our AIP model where $\nu \approx 1.30$ and near that of RP ($\nu_{RP}=4/3$), we therefore expect LRC to become relevant in the vicinity $a<3/2$. In terms of correlations described by Eq.\ref{eq:hurstCorr}, where $-a = 2H$ this leads to the condition $-H<1/\nu$ which we will consider.

Due to its computational efficiency, we use the Fast Fourier transform(FFT) filter technique and provide details in Appendix \ref{appendix:FF Correlation Method}.

This gives our relationship between the Hurst exponent and the appropriate Fourier power spectrum filter function exponent.
\begin{equation}
\beta = 2(H+1)
\label{eq:betaHurstRelation}
\end{equation}

Because the Hurst parameterization is typically 1-d (given by \eqref{eq:hurstExp}), but we rely on a 2d fourier transform parameterized in terms of $\beta$, whose value is shifted by 1 in 2-d relative to 1-d, we need to shift the value of the exponent of $H$ by 1 as is show in \eqref{eq:betaHurstRelation}. 
Thus, if $H=-1.0$ we get no long range correlations, and if $\alpha=0$ we get Brownian long range correlations, which behaves as $k^{-2}$. Since we construct the correlated lattice by applying a Fourier filter characterized by $\beta=2(H+1)$, we use $H$  in range $[-1,0]$.

More importantly, for these reasons we adopt a modified Hurst parameterization which shifts its values by $-1$ sharing the convention of \cite{schrenk2013percolation}. This comes at some risk since in much of the literature use the standard range, $[0,1]$. We choose our parameterization in order to make explicit the need for a mapping between 1-d and 2-d Hurst characterizations. Over the shifted range we preserve the fractal structure of our clusters. We find compact clusters begin forming for $H>0.5$ which corresponds to an unshifted value of $3/2$. Such a value certainly would drive clusters to become compact.   

\section{Static Network Properties}
\label{section:SNP}
In a previous study we characterized some of the essential network properties of our model \cite{ortez2021universality}. This study utilized free edge boundary(FEB) conditions along both axes primarily due to ease of implementation. In a subsequent study we implemented periodic edge boundary(PEB) conditions in order to better establish the universality class of the exponents characterizing the model. We found PEB conditions reliably yielded the infinite lattice limit for the scaling exponents. Finally, AIP's growth algorithm with PEB complements the implementation of site correlations using the Fourier filter technique since FFT's also impose PEB condtions. Here, we outline some of the static network properties and how these change as a result of the input long range correlations.

\begin{figure} 
	\includegraphics[width=0.5\textwidth]{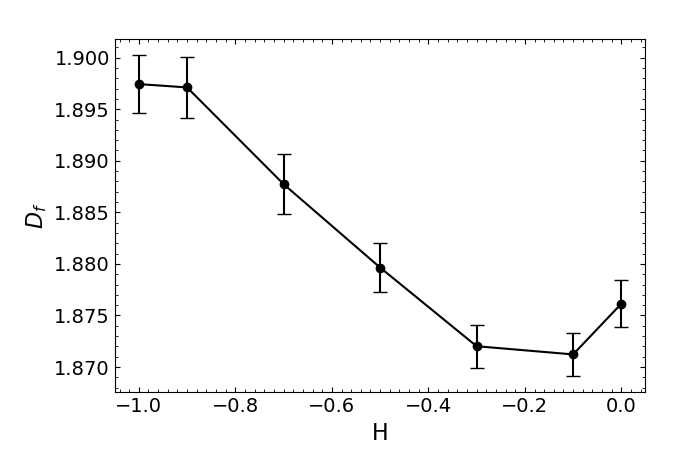}
	\caption{\footnotesize The fractal dimension, $D_f$ for different $H$. For the random case, $D_f = 1.895 \pm 0.016$ which is similar to the expected value of RP. The values all seem to be consistent with one another and doesn't suggest much change as the correlation changes over the range of the study. For $H=0.0$, $D_f= 1.939 \pm 0.028$, which is inconsistent at the $1-\sigma$ from some of the other values.}
	\label{fig:DfHStudy}
\end{figure}

The first characteristic exponent is the scaling of occupied cluster sites, $M(L)$, with lattice size $L$. This scales with characteristic fractal dimension $D_f$ according to:
\begin{equation}
M(L) = L^{D_f}
\label{eq:fractalDim}
\end{equation}
We can easily extract exponent $D_f$ using the well known box counting technique \cite{turcotte1997fractals} and perform linear fit using linear least squares (LLS) on a log-log plot. 

Figure \ref{fig:DfHStudy} shows the extracted $D_f$ for different $H$. For the random case, $H=-1.0$, we reproduce the fractal dimension consistent with RP, $D_f=1.895 \pm 0.016$. We find that input site correlations do not significantly affect the fractal dimension measure in the range of our study. This highlights the macro nature of this measure which is relatively insensitive to changes.

This is weakly consistent with \cite{schrenk2013percolation} which looked at RP with the same long-range correlations and found no change to $D_f$ except for $H>-0.3$ and where $D_f \rightarrow 1.95$ as $H \rightarrow 0$. The authors of \cite{sahimi1996scaling} found similar behavior. Other authors report no detectable change in $D_f$ \cite{prakash1992structural} which considered equivalent $H$ correlation in the range $[-1,0]$. That we observe a change in $D_f$ for $H>-0.9$ illustrates a difference between IP and RP growth mechanisms. 

Perhaps more important is that we observe clear evidence that the site correlations change the density of the invaded sites, since site density is determined by, $\rho \sim L^{d-D_f}$. As observed in the Ising and percolation critical transition, changes in the order parameter induce changes in the density. In our previous characterization of critical behavior of our model \cite{Ortez2022critical}, we showed that because $\rho$ did not change, no suitable notion of an order parameter existed. However, although the change in density reflected by $D_f$ is small, it motivates that LRC should affect the critical behavior of the model. 

Previous studies on the trapping variant of long-range correlated IP in 2D found cluster behavior becomes non-fractal(compact) for $H>0.5$ \cite{knackstedt2002nonuniversality}, though in this study they considered $0 \geq H \geq 1$. In another study the authors of \cite{Vidales_1996} considered a non trapping variant similar to ours and found a minima as we did in the range of our study. While $D_f$ for RP seems to remain unchanged at least for $H<-0.3$, for IP $D_f$ decreases to a minima before likely increasing towards 2 as H increases above zero.  

\begin{figure} 
	\includegraphics[width=0.5\textwidth]{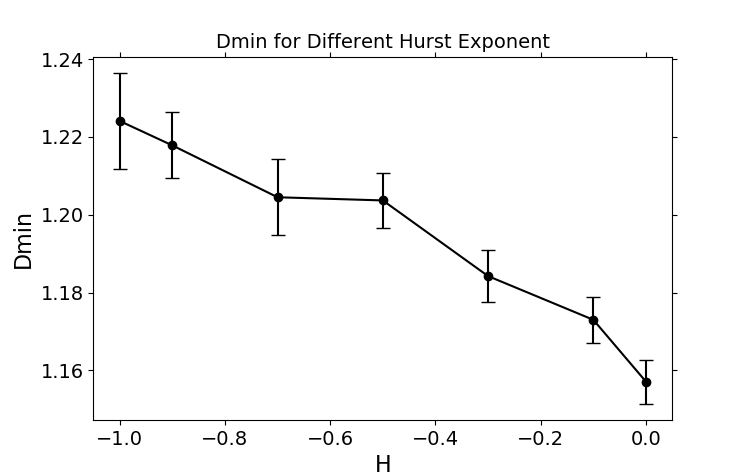}
	\caption{\footnotesize The scaling of distance between sites for different $H$. For the random case $D_{min}\approxeq 1.22$, this tends to decrease as $H$ tends to $0$. The loopless condition will prevent a cluster from becoming compact and $D_{min}$ from becoming 1.}
	\label{fig:DminStudy}
\end{figure}

Though the effect of correlations on $D_f$ is relatively small, we can better understand the effect of correlations on the resulting clusters by looking at the minimum distance between invaded sites. This distance is characterized by scaling exponent $D_{min}$, and it changes more significantly for different $H$. This follows another power law:
\begin{equation}
M\left(l\right) \sim l^{D_{l}}
\label{eq:chemDim}
\end{equation}

Where $M(l)$ is the number of sites within lattice spacing $l$ and $D_l$ is the chemical dimension\cite{havlin1984topological}.
With backbone studies one must be more careful with how boundary conditions are imposed (periodic etc.).  Thus it is preferable to use $D_{l}$ which is largely independent of such affects. Further, what we are really interested in is characterizing the compactness of a cluster which describes the types of paths connecting sites. 
We can relate the Pythagorean distance $r$ and $l$ as:

\begin{equation}
l \sim r^{D_{min}}
\label{eq:pythDist}
\end{equation}

Therefore if $d$ is the path distance from the origin to the boundary of lattice size $L$, then $L=nl$ and by Eq. \ref{eq:pythDist} we can write:
\begin{equation}
d \sim r^{D_{min}}
\label{eq:pathDmin}
\end{equation}

Where $D_{min}$ is the fractal dimension of the shortest path.

We find that as $H$ increases, $D_{min}$ tends to decrease. This behavior is reflected in Figure \ref{fig:DminStudy}. We understand this behavior as follows: for the random case, we expect to find “holes”(trapped regions in IP cluster with loops) in the cluster which are also scale invariant. Paths and the distance between sites in the cluster will necessarily become circuitous. If site strengths are correlated such that similar strengths group together, and given that IP grows by breaking the weakest sites, the IP algorithm will naturally seek out connected regions of weaker sites. This means that fewer portions of the lattice will need to be sampled as the path between two connected sites becomes more direct since it is the result of correlations to create connected regions of weak site regions. Similarly, there will be larger regions devoid of any cluster growth as strong sites will likewise preferentially occupy these regions. This helps us understand the behavior of $D_f$ which is related to the density exponent according to $D_f - 2$. The smaller $D_f$ corresponds to a less dense cluster occupying the lattice, although locally in regions around the cluster, the cluster becomes more dense. This trend starts to reverse for $H>-0.1$, as the dense local cluster regions make up more of the lattice than the large voids filled with strong sites. 

\begin{figure*}[t]
	\includegraphics[width=1.0\textwidth]{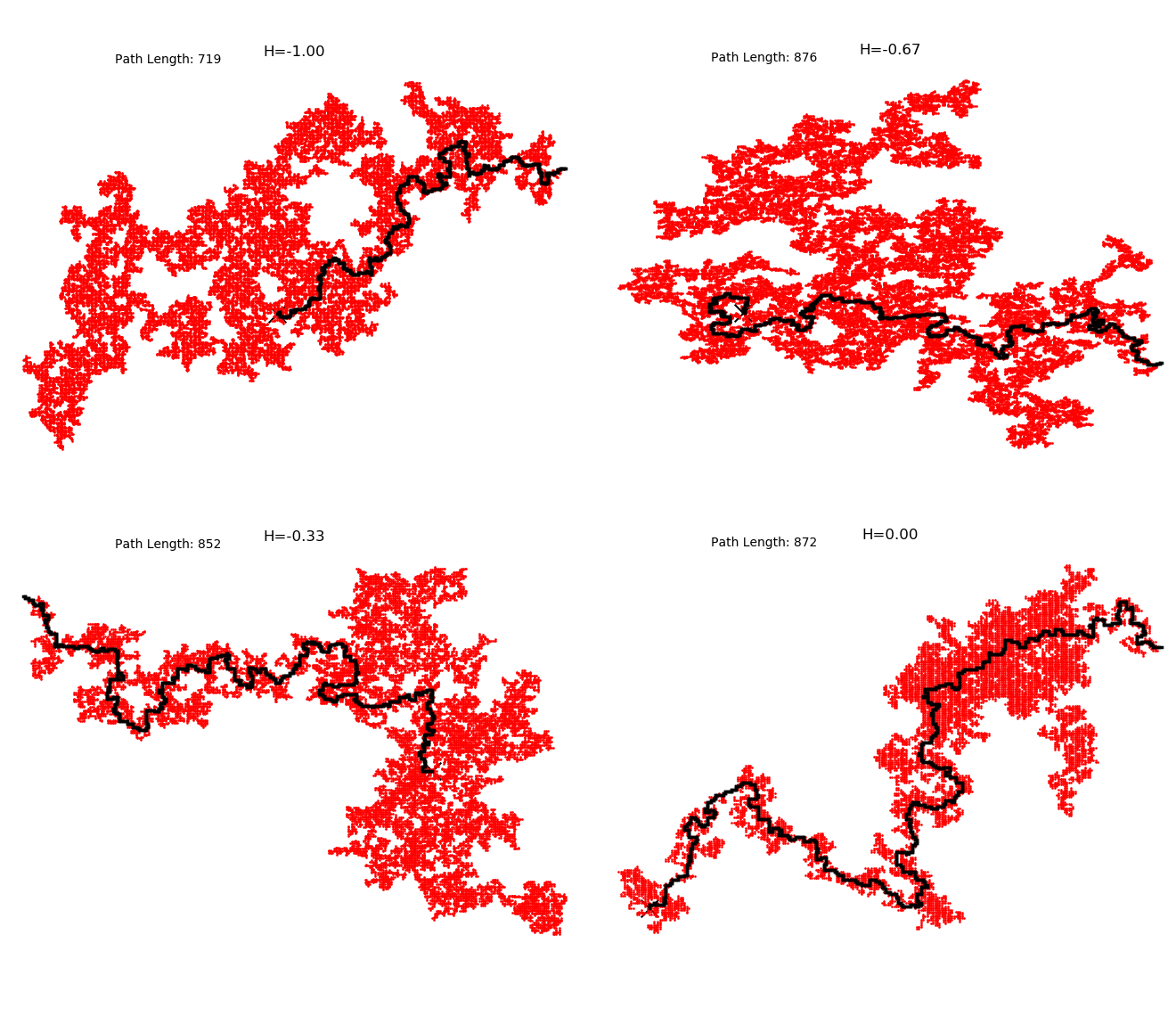}
	\caption{\footnotesize Comparison of clusters grown with different correlation exponent, $H$. As $H \rightarrow 0$ the clusters becomes more dendritic and compact.}
	\label{fig:CorrExample}
\end{figure*}

This behavior is similarly summarized by looking at the backbone exponent $D_{BB}$ as the authors of \cite{prakash1992structural} did with RP. They found that as $H$ increases $D_{BB}$ approaches $D_f$, meaning that the majority of the cluster exists along the cluster backbone. This qualitatively has the effect of causing the  cluster to become both more dendritic and compact as the Hurst exponent increases. This is shown in Figure \ref{fig:CorrExample}.

\section{Critical Threshold}
\label{section:critThresh}
One of the most important features of percolation is its relation to critical phenomena \cite{stephen1976percolation}. In the previous section we characterized the static network properties of the entire AIP cluster, however, criticality is characterized by the structure of fluctuations near the critical point. In this section, we show how the critical threshold of our model, $T_c$, changes under the application of long-range correlations to lattice site strengths. Previous studies with RP on long-range correlated lattices have shown that the $p_c$ changes depending on the Hurst parameter, $H$ \cite{prakash1992structural}. Other authors used $p-p_c \sim L^{-1/\nu}$ relationship to determine $p_c$, but this becomes problematic since $\nu$ changes as a result of long range correlations in a non-trivial way \cite{schrenk2013percolation}.
 
We begin by looking at the distribution of site strengths of the invaded cluster. In IP all lattice sites are randomly assigned values from a uniform distribution in the range [0,1], but when looking at the distribution of the strengths of invaded sites, we find the selection of strengths to be a regular subset of assigned strengths. In particular, in the limit where the number of invaded sites,$N$, becomes infinite, the invaded strength distribution is described by a step function:

\begin{displaymath}
\lim_{N \to \infty} p(r) = \left\{
\begin{array}{lr}
k & 0 \ge r \ge r_{max} \\
0 & r > r_{max}
\end{array}
\right.
\end{displaymath}

where a random strength, $r$, has constant probability $k$, of being invaded up to some strength, $r_{max}$. These are related according to $1/k = r_{max}$, and its been shown that $r_{max}=p_c$ where $p_c$ is RP's critical occupation probability \cite{chayes1985stochastic}.


\begin{figure} 
	\includegraphics[width=0.5\textwidth]{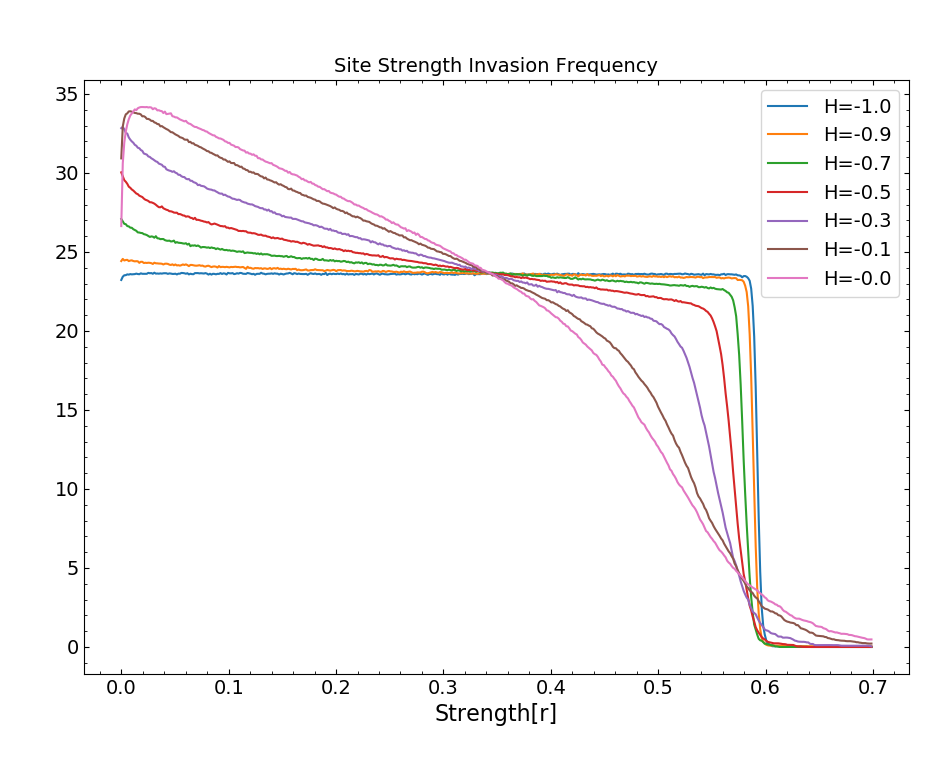}
	\caption{\footnotesize The changing distribution of invaded strengths for different correlation Hurst parameter, $H$. For the independent random case($H=-1.0$) we recover an approximate step function reflecting constant probability of invading a particular site up to $r_{max}$ anywhere in the cluster. As spatial correlations increase it becomes increasingly likely to sample weaker sites.  }
	\label{fig:invDist}
\end{figure}

A flat uniform distribution of invaded sites is evidence that regardless of where in the lattice the growth takes place, the likelihood of a particular strength to be invaded remains constant.
If instead we could sample weaker sites with more regularity than stronger ones, we would no longer observe a flat probability, and subsequently, the threshold would change depending on the local ratio of weak/strong bonds. This is precisely the scenario introduced when introducing long range correlations into the assigned strengths. Fig \ref{fig:invDist} shows how the distribution of invaded sites changes as a result of changing correlation exponent, $H$. 

\begin{figure} 
	\includegraphics[width=0.5\textwidth]{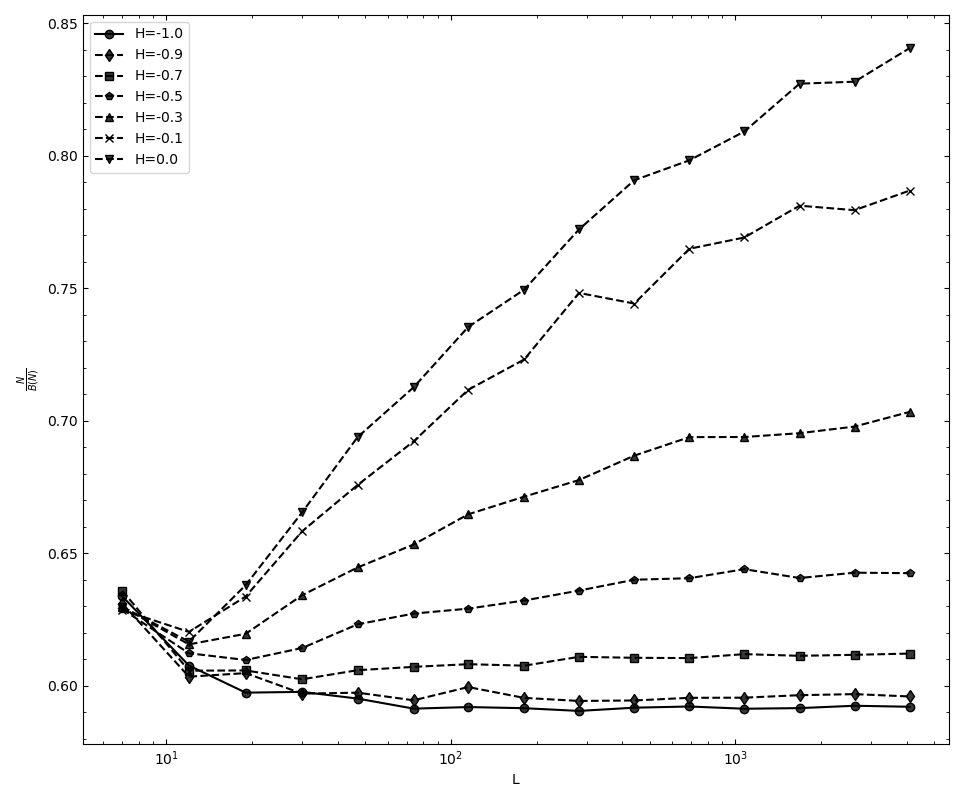}
	\caption{\footnotesize  Here we show how the "bulk to boundary" ratio changes as a function Hurst correlation exponent $H$. For random case($H=-1.0$) we see the ratio approach $p_c=T_c$, but for $H>-0.5$ the ratio fails to asymptote to a particular value.  }
	\label{fig:B2Bstudy}
\end{figure}
These changes to the strength distribution introduce the following feature: the local strength environments produces sufficiently different thresholds such that the notion of global lattice threshold breaks down. 
\begin{table*}[t]
	\begin{center}
		\setlength{\tabcolsep}{10pt}
		\begin{tabular}{l c c c} \hline \hline
			& $D_{f}$ & $D_s$ & $2-\eta$  \\
			\hline
			$H=-1.0$ & $1.897 \pm 0.003$ & $1.860 \pm .002$ & $1.804 \pm .009$  \\
			$H = -0.9$ & $1.894 \pm 0.003$ & $1.857 \pm .002$ & $1.79 \pm .01$ \\
			$H = -0.7$ & $1.888 \pm 0.003$ & $1.863 \pm 0.002$ & $1.79 \pm .01$\\
			$H = -0.5$ & $1.880 \pm 0.002$ & $1.867 \pm 0.001$ & $1.794 \pm 0.007$ \\
			$H = -0.3$ & $1.872 \pm 0.002$  & $1.855 \pm 0.002$ & $1.78 \pm 0.01$ \\
			$H = -0.1$ & $1.871 \pm 0.002$  & $1.850 \pm 0.002$ & $1.77 \pm 0.01$ \\
			\hline \hline
		\end{tabular}
		\caption{ Static scaling exponents.}
		\label{table:static_exponents}
	\end{center}
\end{table*}
Input correlations of type in Eq. \ref{eq:betaHurstRelation} will produce produce mean strength fluctuations defined as  are described by described by:
\begin{equation}
<u(r^\prime)u(r^\prime+r)>=\left\langle\delta s^2\right\rangle - \left\langle\delta s\right\rangle^2
\end{equation}

where $\delta s = u_i - x$ and $u_i$ is strength of the $i$th site and x is the random non-correlated component of the strength. We find the mean strength fluctuations are also described by:
\begin{equation}
\left\langle\delta s^2\right\rangle - \left\langle\delta s\right\rangle^2 \sim r^{-2H}
\end{equation}
which we recognize as also describing the second moment of the strength distribution, which will have well-defined mean for $2H>2$ and well-defined variance for $2H>3$. Thus, by construction, the variance of average strengths is poorly defined since the tail events are not exponentially bounded. This results in infinite variance. Moreover, even average values for quantities resulting from averaging over distinct regions will not be well behaved. Therefore, any averaged macroscopic quantity will be poorly behaved. 

\begin{figure*}[t] 
	\includegraphics[width=1.0\textwidth]{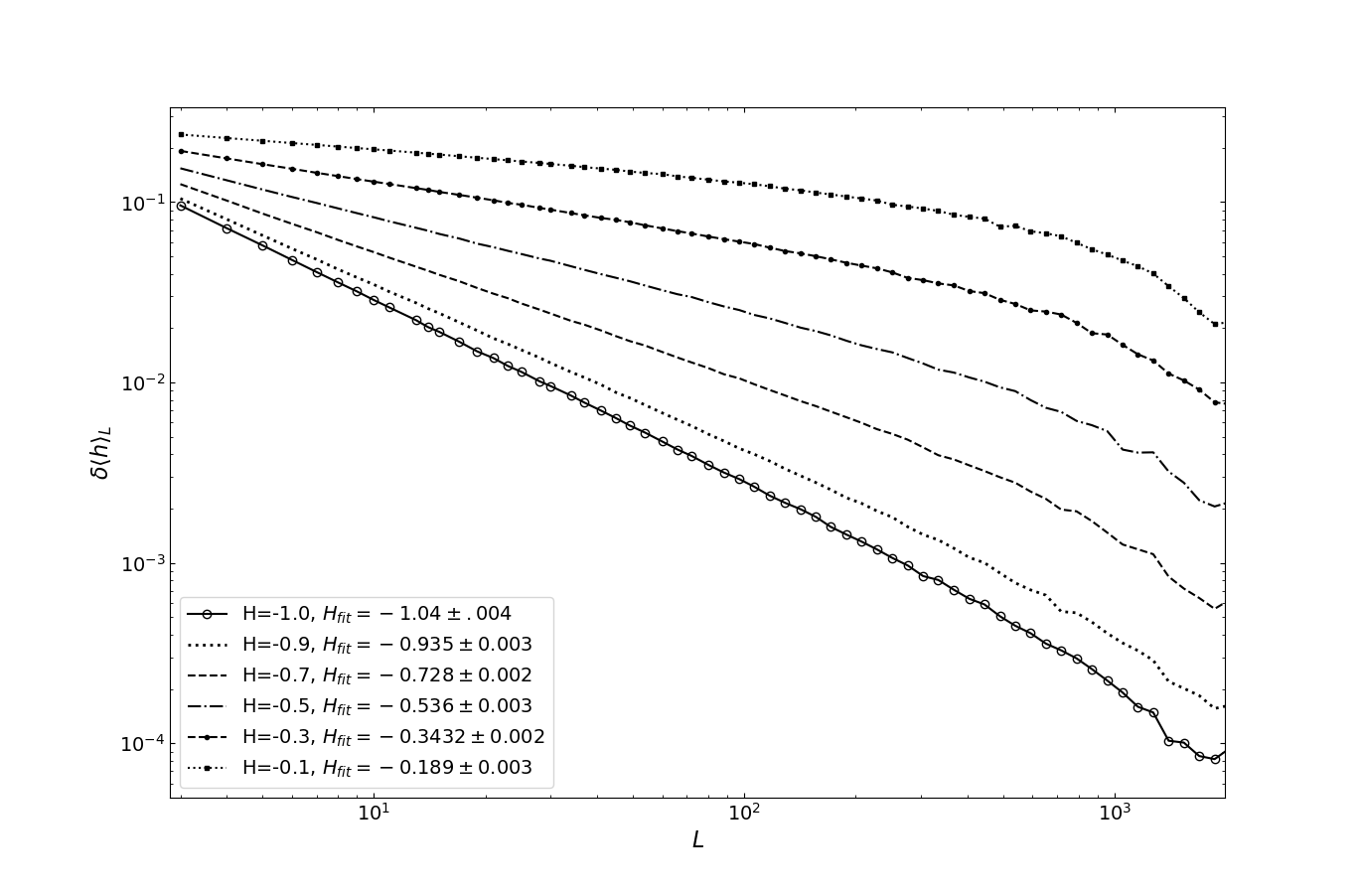}
	\caption{\footnotesize  Mean site strength fluctuations. We show the expected scaling of lattice site strength fluctuations, $\delta \langle h \rangle_L \sim L^{-H}$. }
	\label{fig:strength_fluctuations}
\end{figure*}

An alternative notion for a burst could rely instead on a "bulk to boundary" ratio, $r_{BB}$. The authors in \cite{prakash1992structural} used a similar argument to determine $p_c$ with long-range correlations where they determined $p_c$ by noting which $p_{occ}$ produced a ratio of 1 between the perimeter of filled and unfilled sites. Using a similar strategy authors argued that using a "bulk to boundary" ratio is a generalized way to determine the critical occupation probability \cite{mertens2017percolation}. However, determining the ratio analytically using:
\begin{equation}
\lim_{N \to \infty} \frac{N}{B(N)}=T_c
\end{equation} 
leads to slightly different results since $T_c \rightarrow p_c$ only in the random case. In Leath's  original paper \cite{leath1976cluster} the expression for the probability of finite clusters of size $n$ with $b$ empty perimeter sites assumed sites with independent random probabilities. One must instead empirically determine the ratio leading to a scale invariant distribution of bursts.  

We empirically determined the ratio, $r_{BB}$ for our clusters for different $H$. We found that the behavior of $r_{BB}$ did not universalize in any way to allow us to preserve the notion of a collective critical point. Not only do the values of stable ratios change, but we find that  for $H>-0.5$, $r_{BB}$ fails to asymptote to a fixed value. These results are shown in Fig \ref{fig:B2Bstudy}.

With random AIP, we established the existence of a critical threshold, but with long-range correlations, these relationships no longer hold. In the next section(sect. \ref{section:crit}), we discuss how the phase transition is smoothed such that there is no longer a power-law divergence with the control parameter as $T \rightarrow T_c$. The notion of criticality itself begins to break down, but its worth wondering whether we have the correct notion of the critical control parameter such that we observe universal critical behavior in the presence of long-range correlations.

\section{Correlated Critical Behavior - $\xi, \nu$}
\label{section:crit}
\begin{figure*}[t]
	\includegraphics[width=1.0\textwidth]{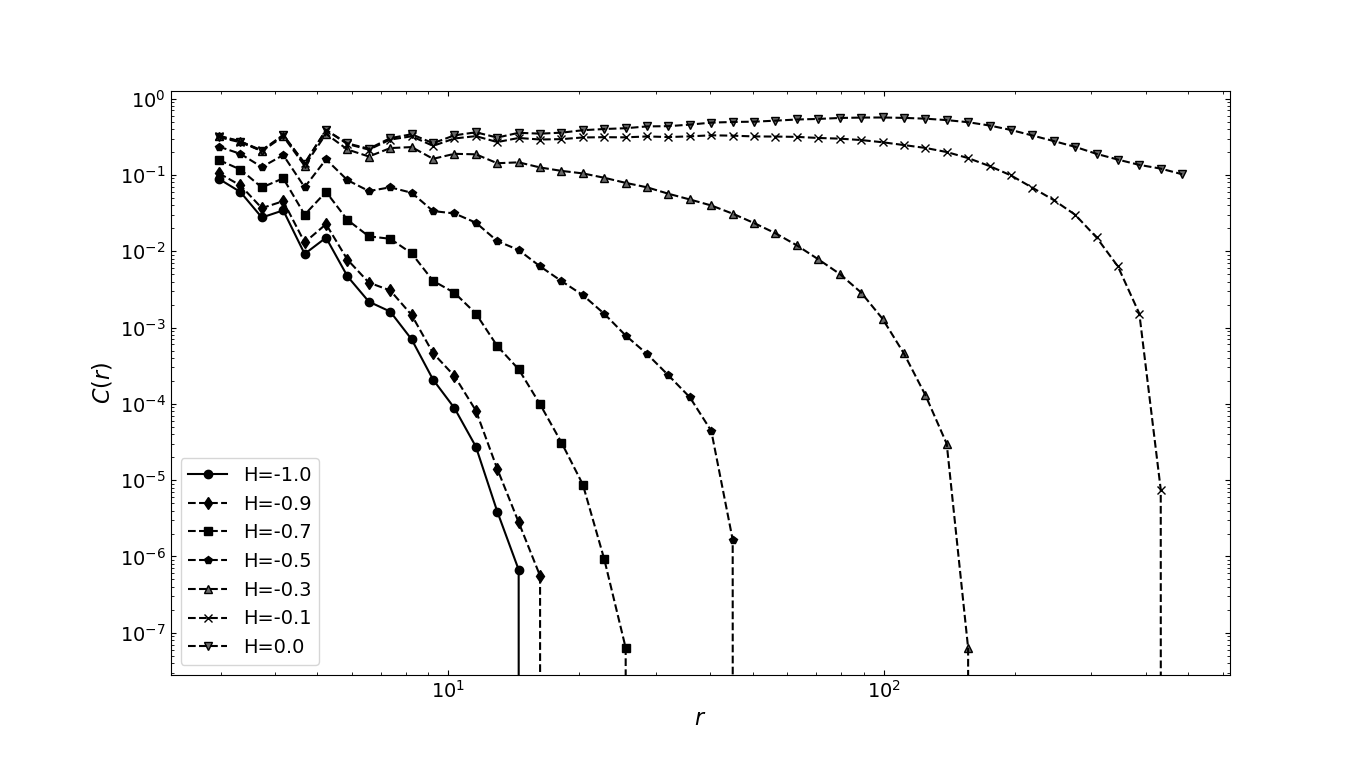}
	\caption{\footnotesize  Here we show how the correlation function for different Hurst correlation values, $H$. We fix the threshold to be $T=0.25$ for all $H$, and we observe how the exponential decay constant, $\xi$ varies from $\sim 1$ in the random case to $\sim L$ in the maximally correlated case.  }
	\label{fig:Cr_All}
\end{figure*}

For RP in the absence of additional lattice site correlations, the behavior of critical fluctuations in average site occupation scale according to $\delta \langle p_{occ}\rangle \sim L^{-1/\nu_{R}}$(where $\nu_R$ is the RP correlation length scaling exponent). This is the emergent structure that allows scale invariant connected burst sequences to form. In the context of AIP, the set of invaded sites and their associated strengths will form a subset of strengths in the range [0, $T_c$](where $T_c=p_c$) with characteristic length $L^{-1/\nu_I}$ ($\nu_I=1.3$ slightly different from RP's $\nu_{R}=4/3$). Thus, bursts grown at $T_c$ reproduce RP's incipient infinite cluster (IIC), and importantly, do so without producing the associated distribution of finite clusters. Therefore, with AIP near $T_c$, we continually sample IIC's subject to an environment of already populated and grown IIC. If we now input additional correlations that yield fluctuations in strength according to $\delta \langle h \rangle \sim L^{-H}$, this alters the mechanism responsible for long-range structure, and this also affects the effective correlation length and burst size distribution, thereby altering the overall critical behavior. We confirm the expected site strength fluctuations resulting from our long-range correlation scheme and is shown in Fig \ref{fig:strength_fluctuations}.
\begin{figure} 
	\includegraphics[width=0.5\textwidth]{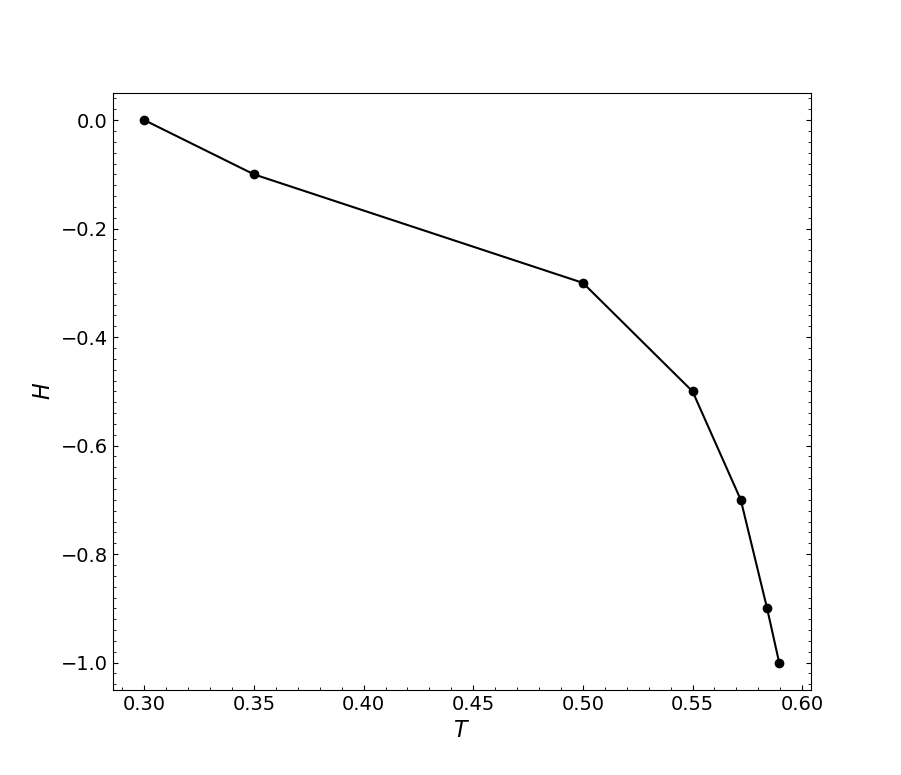}
	\caption{\footnotesize  Here we show how the value of thresholds,$T$, where $\xi(T,H) \sim L_{sys}$ giving rise to scale invariant burst distribution.}
	\label{fig:TcvsH}
\end{figure}


We begin by addressing the question of how site strength correlations affect the correlation length, $\xi$. This choice is motivated by fundamental finite size scaling hypothesis, the bedrock of criticality. 
For lattice systems, the notion of correlation length is generally understood by the correlation function (pairwise correlation function) which empirically has been established to behave according to
\begin{equation}
C(r) \sim r^{d-2+\eta}e^{-r/\xi}
\label{eq:pairwiseCorrFunc}
\end{equation} 
Therefore, the correlation length $\xi$ characterizes when random correlations become exponentially suppressed as a function of distance, $r$. If $\xi \sim L_{sys}$, then the correlations display long range behavior descried with power law $C(r) \sim r^{2-d+\eta}$. 
\begin{figure*}[t] 
	\centering
	\includegraphics[width=1\textwidth]{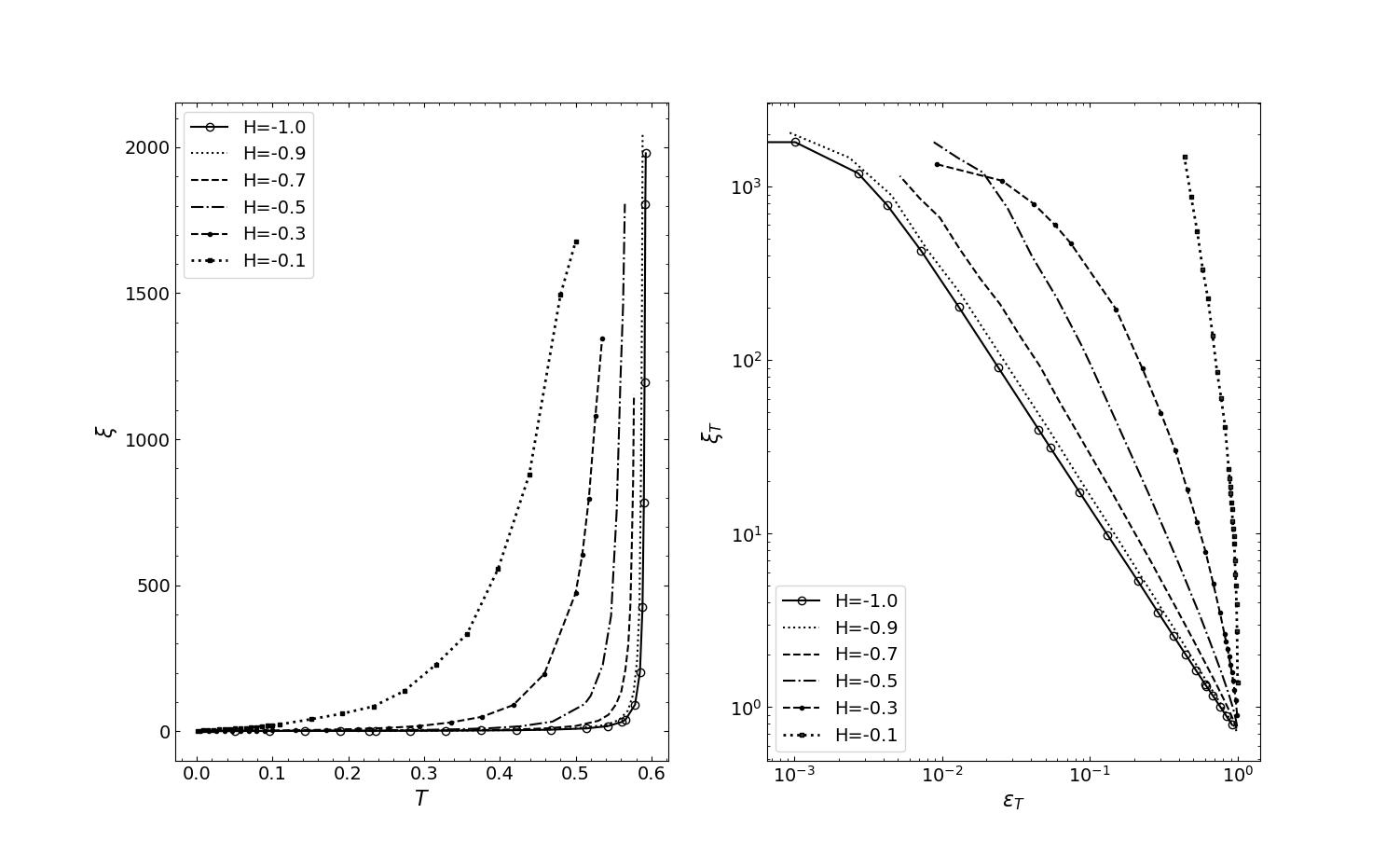}
	\caption{\footnotesize Correlation length comparison for different Hurst correlations, $H$. Generated from the statistics of over $10^7$ bursts grown with PBC on lattice of size 4096x4096. (left) We plot the correlation $\xi_T$ vs burst threshold, $T$. (right) We plot the burst critical scaling with critical parameter, $\epsilon_T$. For the random case with $H=-1.0$, we get correlation length scaling exponent, $\nu=1.30$, which is nearly similar to the RP value. Also, we can confirm that for $H>-3/4$ we get correlation length scaling exponent given approximately by $\nu_H \sim 1/H$.  }
	\label{fig:corrLen_allH}
\end{figure*}

Though the characterization of $C(r)$ provides a rather simple, intuitive understanding, it is seldom used in the literature since the pairwise correlation function is often very cumbersome to calculate. Its computations scale according to $O(N^2)$ with $N$ sites/particle. Given an individual cluster ensemble contains $10^7$ sites, of which we use $10^2-10^3$ ensemble elements to obtain reliable statistics, aquiring the requisite statistics quickly becomes computationally prohibitive. Fortunately, we were able to rely on a highly optimized and parallelized implementation to efficiently compute, $C(r)$ \cite{2020MNRAS} which gave good results.

We find that the effect of long range correlations on burst formation is to change the required threshold that is likely to produce a scale invariant burst (when its  $\xi \sim L_{sys}$). This effect of $H$ on the correlation function, $C(r)$ is shown in Fig \ref{fig:Cr_All}. In order to establish $H$ dependence, we fixed $\epsilon_{T}$ for all curves and vary $H$.  In the random case, we observe the usual $\xi(T)$ dependence that becomes exponentially suppressed for burst sizes of order 2, but with correlations we observe the additional $H$ dependence, $\xi(T,H)$ which can greatly extend the correlation length despite keeping $\epsilon_T$ fixed. In fact, we can nearly reproduce the $C(r)$ near the critical value by merely changing $H$.  The expected relation becomes,
\begin{equation}
\xi(T,H) \sim \epsilon_{T}^{1/H}
\label{eq:Corr_len_T_H}
\end{equation}
which agrees with the empirical results of Fig \ref{fig:Cr_All}. In our formulation, the random case corresponds to $H=-1.0$ which yields a correlation length of order $1/\epsilon_{T}$, while for maximally correlated case, $H=0$, yields a strongly diverging correlation length and becomes limited by the system lattice size, $L_{sys}$. 


As Hurst site correlations become increasingly dominant ($H\rightarrow0$), there exists crossover behavior where it becomes the dominant scaling mechanism.
We therefore define characteristic lengths, $\xi_H$ and $\xi_b$ to correspond to these respective length scales. Briefly, we can anticipate the behavior following the crossover by recalling the arguments of the extended Harris condition (briefly derived in Appendix \ref{appedix:LRC}), where for longest range dependencies the slowest vanishing mechanism will dominate. But what is also of interest is the behavior where multiple correlation mechanisms compete.
\begin{figure*}[t] 
	\includegraphics[width=1.0\textwidth]{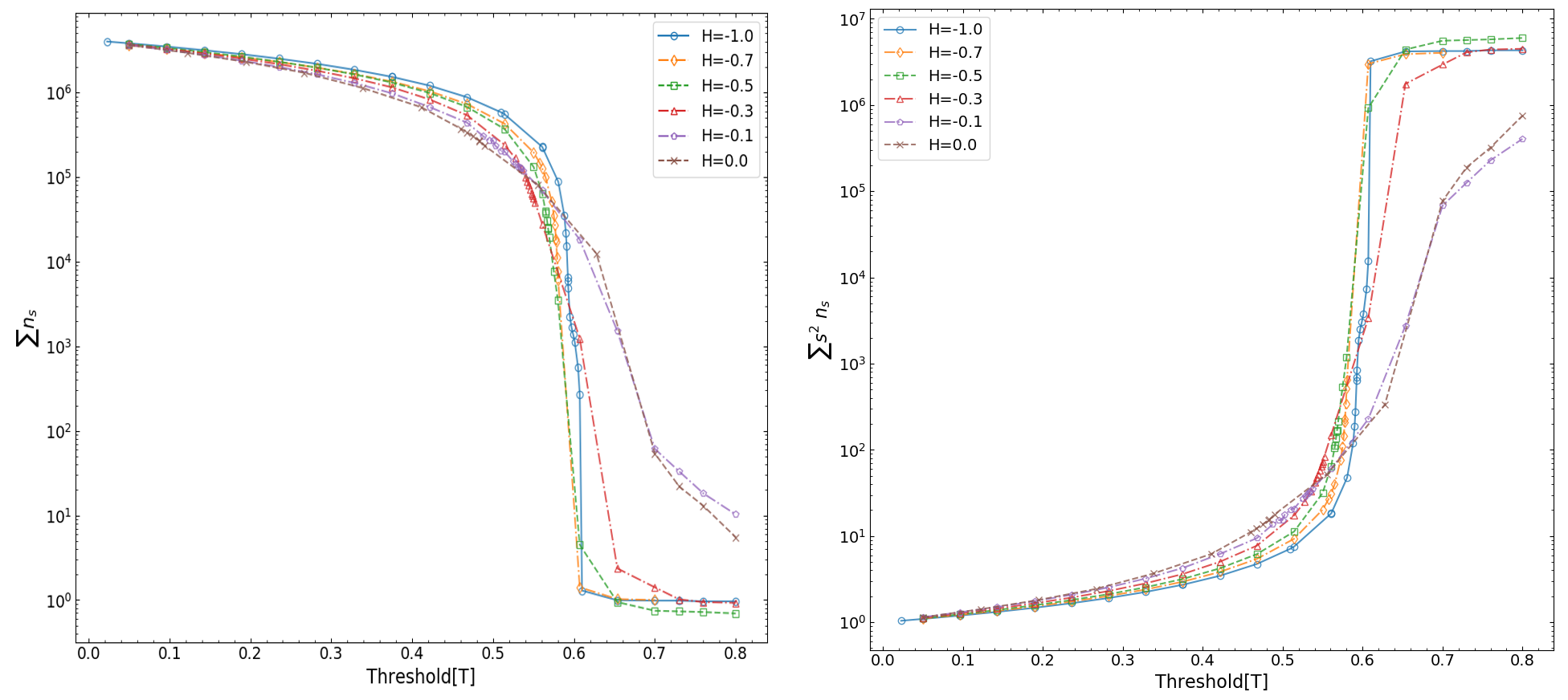}
	\caption{\footnotesize  Here we show how the cluster distribution moments fail to diverge, except for the random case ($H=-1.0$). The transition region becomes increasingly broadened as the Hurst correlation parameter increases. (left) Represents the zeroth moment or naturally corresponds to the number of bursts as a function of threshold, $T$. (right) The behavior of the second moment corresponding to the average burst size as function of threshold, $T$. }
	\label{fig:critMoments}
\end{figure*}

We expect from the extended Harris criteria that the site strength correlations become relevant when their associated correlation scaling becomes larger than that of the random case, $\nu_H=1/H>-4/3$. While we do see some minor affects for $H=-0.9$ on the correlation length scaling, we generally observe behavior consistent with the extended Harris criteria, which tells us that for $H>-3/4$ the correlation exponent, $\nu_H$ is given by,
\begin{equation}
\nu_H = 1/H
\label{eq:nu_H}
\end{equation}
We confirm this behavior by calculating $\xi_H$ in the standard way \cite{Ortez2022critical,stauffer1994introduction}. The obtained scaling is reported in Table \ref{table:crit_exponents} and shown in Fig \ref{fig:corrLen_allH}.



\begin{figure}
	\centering
	\includegraphics[width=0.5\textwidth]{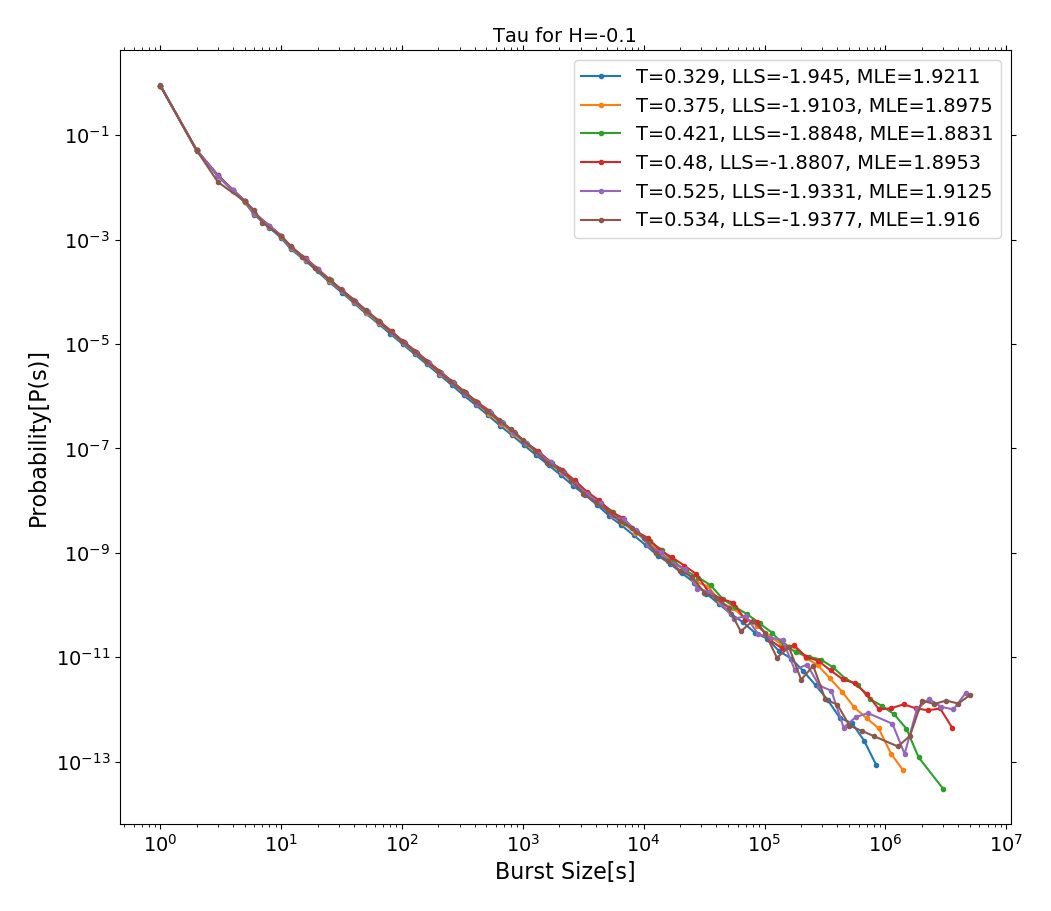}
	\caption{\footnotesize Burst scaling for correlated case, $H=-0.10$. Each curve represents the burst statistics for different burst thresholds, $T$. Also shown are the linear fits to each curve represented by LLS and by MLE methods. The threshold becomes degenerate as a wide range of threshold lead to similar burst scaling statistics. We see a family power-laws for thresholds in the range $[0.329,0.534]$ which produce scaling exponents $\tau$ in the range $[1.88,1.92]$}
	\label{fig:corrIP_degTau}
\end{figure}

\section{Correlated Critical Behavior - $n_s(\tau, \sigma)$}
Following what was shown in Section \ref{section:critThresh} where the critical point becomes increasingly degenerate for $H \rightarrow 0$, the associated critical quantities also smear over an increasing range of thresholds.
We can observe this regime in Fig \ref{fig:critMoments}, where we can see similar behavior of the burst distribution moments (1st and 2nd) up until the vicinity of the critical point where the behavior of the curves depart. The width of the critical transition region scale with $\xi_H^{-H}$, and the fluctuations become dominated by the Hurst correlation statistics. Further, we can notice that the curves of Fig \ref{fig:critMoments} representing different $H$ do not lie on top of each other, indicating that the burst distribution scaling exponents ($\tau, \sigma$) ought to be different.


Even in the regime where Hurst correlations dominate ($\xi_b<\xi_H$), we observe the existence of scale free burst distributions for all ranges of $H$ we considered. This means that for each $H$ we get unique critical scaling behavior. As was done previously \cite{Ortez2022critical}, the critical behavior is largely characterizing by the burst distribution $n_s(\tau, \sigma)$. Thus, we define the usual control parameter $\epsilon_T = (T_c-T)/T_c$ where because the minimum threshold required to grow an IIC changes with $H$ (shown in Fig \ref{fig:TcvsH}), we need to account for the dependence of $T_c(H)$. Also, we need to insure condition $l < \xi_b$ which is imposed by requiring $\epsilon_{T} > \xi_b^{-1/\nu_H}$. 

Of course one of the nice features of working with the burst distribution is that we can directly calculate the expected behavior of the average burst size scaling. This is done with the usual moment calculation, 
\begin{equation}
M_k =\epsilon_T^{\frac{1+k-\tau}{\sigma}} \int_{0}^{1}dz\ z^{k-\tau}f[z]
\label{eq:momentInt}
\end{equation}
where here $k=2$, $z=(T_c-T)s^\sigma$, and in the upper limit, we observe the relation $s \gg (T_c -T)^{-1/\sigma}$. Thus, $s_\xi(T) = (T_c -T)^{-1/\sigma}$ behaves as the exponential cutoff cluster size for the cluster size distribution. Since, the integrand evaluates to a constant, we once again get the familiar $\gamma$ exponent relationship, 
\begin{equation}
\gamma = \frac{2-\tau}{\sigma}
\label{eq:gamma_H}
\end{equation}
However, we must characterize the burst cutoff size $s_\xi$, where we observe length burst size relation, $s_\xi = \xi^{D_s}$ from Eq.\ref{eq:fractalDim} and leads to the exponent relation,
\begin{equation}
\frac{1}{\sigma} = D_s \nu
\label{eq:sigma_H}
\end{equation}
There is a slight distinction between $D_s\approx 1.865$ and $D_f\approx 1.896$ in the random case, which does not significantly affect the critical behavior, and these differences decrease for increasing Hurst correlations since $D_s \sim D_F$. As was discussed previously and found in Section \ref{section:SNP}, fundamental mass-length scaling exponents $D_F,D_s$  change very little for changing Hurst correlations, therefore, this leads to the important behavior of $\nu$ which we introduced in the previous section and found that it was heavily dependent on $H$.

Therefore, in light of Eq. \ref{eq:sigma_H} and since the measure of $\nu$ is more reliable (than that of $\sigma$ shown in \cite{Ortez2022critical}), we combine Eq.\ref{eq:sigma_H} into Eq.\ref{eq:gamma_H} and establish the following relation, 
\begin{equation}
\gamma = (2-\tau)D_s \nu
\label{eq:gamma_nu}
\end{equation}
which will allow us to determine how the average burst size ought to behave as a function of exponents $\tau, \nu$. The results are shown in Table \ref{table:crit_exponents}. This forms the basis for a family of interdependent critical exponents and most importantly the existence of a scale invariant burst distributions characterized by exponents, $\tau$, $\sigma(\nu)$. Some exponents change very little ($D_f, D_s, 2-\eta$), while others change quite noticeably ($\tau,\nu$). These family of exponents potentially give rise to a whole host of distinct universality classes if these properties are extrapolated to the entirety of a system, but we are careful to note that these relations only hold up to a certain length scale, namely, for $l < \xi_H$.

\begin{table*}[t]
	\begin{center}
		\setlength{\tabcolsep}{10pt}
		\begin{tabular}{l c c c c c } \hline \hline
			& $T_c$ & $\tau$ & $\nu$ &$\gamma_{obs}$ & $\gamma_{th}$ \\
			\hline
			$H=-1.0$ &$0.5926(5)$ & $1.594(2)$ & $1.301(2)$ &$0.971(5)$ &$0.9825$  \\
			$H = -0.9$ & $0.590(1)$ & $1.597(2)$ & $1.359(2)$ &$1.025(3)$ & $1.017$\\
			$H = -0.7$ & $0.580(5)$ & $1.635(2)$ & $1.47(2)$ &$1.041(5)$ & $1.00$\\
			$H = -0.5$ & $0.570(5)$& $1.711(2)$ & $1.95(2)$ & $1.066(3)$  & $1.052$ \\
			$H = -0.3$ &$0.54(1)$ & $1.810(2)$  & $3.1(1)$ & $1.10(2)$ & $1.09$ \\
			$H = -0.1$ & $0.48(9)$  & $1.90(5)$ & $7.6(5)$ & $1.18(1)$ & $1.4$ \\
			\hline \hline
		\end{tabular}
		\caption{ Critical scaling exponents. comparison of scaling exponents for AIP model with different Hurst correlations $H$, where $H=-1.0$ is the random case($H=0$ in the usual formulation)and Hurst correlations increase with increasing $H$. We used a 4096x4096 lattice with PBC to generate statistics. In order to account for any remaining finite size affects, we set the burst size threshold to be $10^6$. We used at least $10^9-10^6$ bursts for all statistics, depending on the proximity to the critical point. The error represented in parenthesis of the final digit is the error in LLS fit. We find that critical relations start breaking down as $H\rightarrow0$, indicating critical processes no longer govern behavior. }
		\label{table:crit_exponents}
	\end{center}
\end{table*}

\section{Fluctuation Dissipation Theorem and $\tau(H)$?}
One might expect a precise relationship governing $\tau(H)$, since from Table  \ref{table:crit_exponents} we can see an increasing magnitude of $\tau$ as $H$ increases. This would be a neat way to summarize the effects of Hurst correlations on critical behavior, given the central role of $\tau$ and $n_s(\tau)$. However, the situation is not quite so simple as solving Eq. \ref{eq:gamma_nu} for $\tau$ leaves non trivial $H$ dependence in both $\gamma$ an $\nu$.   
A complementary technique for calculating the average burst size comes from the fundamental, fluctuation dissipation theorem \cite{ma2018modern} which relates the susceptibility to average site strength correlations. This provides another relation for $\gamma(H)$ which in conjunction with Eq. \ref{eq:gamma_nu} ought to allow one to determine the relation for $\tau(H)$.  The fluctuation dissipation theorem in the language of percolation becomes a relation between the pairwise correlation function mentioned previously, $C(r)$, and the average burst size, $\langle s \rangle$ \cite{coniglio1979cluster, stauffer1994introduction}. The relationship is given by,
\begin{equation}
\langle s \rangle = 1/V \int dV C(r)
\label{eq:fluctDissThm}
\end{equation}
where $V$ is typically taken to be the correlation volume given by, $\xi^d$. Again, we compute $C(r)$ using \cite{2020MNRAS} with good results. With $C(r)$ given by Eq.\ref{eq:pairwiseCorrFunc} we can rewrite Eq.\ref{eq:fluctDissThm} as,
\begin{equation}
\begin{split}
\langle s \rangle &= \xi^{-2} \int dr \ r^{2-\eta}\exp^{-r/\xi} \\
& = \xi^{-2} \xi^{3-\eta} \int_0^{\infty}dz \ z^{3-\eta} \exp^{-z} \\
& \sim \xi^{1-\eta} \\
& \sim \epsilon_{T}^{\nu(1-\eta)}\\
\end{split}
\label{eq:critSuscScaling}
\end{equation}
where $z=r/\xi$ and the integrand yields a constant. The last line also makes use of the usual critical scaling relation $\xi \sim \epsilon_{T}^{-\nu}$ which we know is valid on smaller length scales. This gives another relation,
\begin{equation} 
1-\eta = (2-\tau)D_s
\end{equation}
by plugging into Eq. \ref{eq:gamma_nu}. 

Since $1-\eta$ and $D_s$ are nearly constant for changing $H$, we would similarly expect $\tau$ to be nearly constant, but this of course is not true as is shown in Table \ref{table:crit_exponents}. Solving the above equation for $\tau$ gives $\tau \approx 1.59$, which is the burst distribution scaling for the random case and for $\gamma \sim 1$. 
 

The failure of the fluctuation dissipation theorem as originally defined provides further evidence that behavior departs from the usual critical fluctuation scaling dominating percolation transitions. Again, the primary complication is the existence of multiple competing correlation lengths, which 1) are not properly accounted for in integration given by Eq. \ref{eq:critSuscScaling} and 2) the failure of a global, uniform lattice transition. The effect of Hurst correlations is to produce distinct regions of site strengths which alter the threshold required to grow scale invariant bursts within these pockets. This range of thresholds is given by $\xi_H ^{-1/H}$ and therefore quickly increase as $H\rightarrow 0$. This fundamentally alters the mechanism generating the critical Fisher distribution from the avalanche burst type to Hurst correlations, and importantly admits mechanisms that are non critical in origin (at least not governed by $\xi \sim \epsilon_{T}^{-\nu}$) to produce a scale invariant burst distribution. This last point is particularly important since the presence and characterization of the Fisher distribution has largely been adequate in motivating and confirming critical behavior. Here, we have an example of such a distributions absent the usual mechanisms that drive a critical transition; one where the $\epsilon$ characterization fails to describe scaling behavior. Still, the correlation length of the system is a power law, but it does not diverge as a function of the proximity to the critical point, and therefore fails to satisfy the requirements of second order phase transition mechanics. 

\begin{figure}
	\centering
	\includegraphics[width=0.5\textwidth]{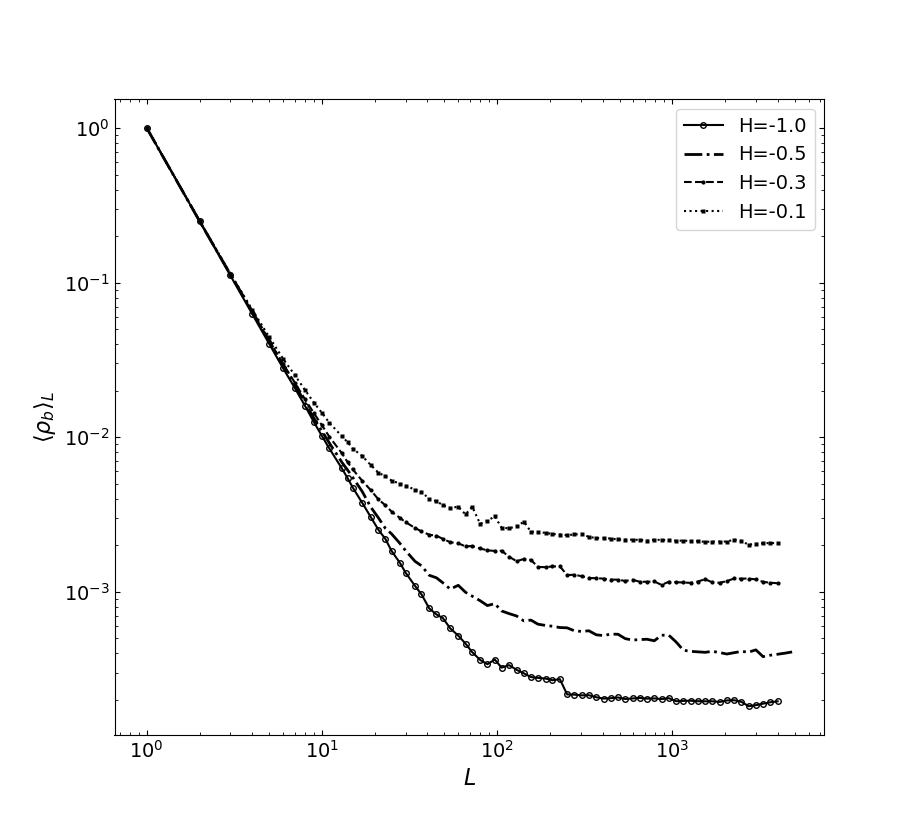}
	\caption{\footnotesize We show the crossover behavior inherent to AIP subject to different, $H$. We find that average burst epicenter density follows correlation length determined by $\xi_b \sim \epsilon_T^{-1/2}$ for small scales and $\xi_H^{-H}$ at the crossover.}
	\label{fig:burstEpicenters}
\end{figure}
\section{Scaling Crossover Behavior and Multifractality}
In our previous work \cite{Ortez2022critical}, we found the burst epicenter scaling was different from the general site mass scaling, $D_f$. This result is significant because it suggests the existence of multiple correlation lengths. One correlation length is associated with the likelihood of sites to occupy different burst clusters, and the other length characterizes the likely distance between burst centers. The average burst epicenters density $\langle \rho_b \rangle$ as a function of length scale exhibited expected crossover behavior for lengths greater than $\xi_b$, where $\langle \rho_b \rangle$ scaling becomes nearly uniform, and for lengths less than $\xi_b$ we observe scale invariant $\langle \rho_b \rangle$ consistent with site density scaling $D_f$. The scaling of $\xi_b$ as function of burst threshold also was found to behave with mean-field correlation exponent, $\xi_b \sim \epsilon_{T}^{-1/2}$, suggesting that burst epicenters were distributed as a random walk about the lattice and is consistent with uniform scaling, $D=2$. 

With the addition of Hurst correlations, there is yet another correlation length to factor, and again we find scale invariant $\langle \rho_b \rangle$ on length scales less than $\xi_b$. However, since the correlation length for different $H$ follow $\xi_H^{-H}$ dependence, the crossover length scale changes for different $H$. Fig \ref{fig:burstEpicenters} shows how the crossover length of $\langle \rho_b \rangle$ changes for different $H$, becoming mean-field and uniform above the crossover. Namely, we find that this length follows $\xi_H^{-H}$ and gets shorter as $H \rightarrow 0$.   

The behavior of $\langle \rho_b \rangle$ provides an instructive way to understand why the burst distribution $n_s(\tau, \sigma)$ changes. The crossover length above which $\langle \rho_b \rangle$ becomes uniform, indicates that burst with characteristic length above this must uniformly spaced. Because the burst densities follow random walk statistics with a denser distribution of burst epicenters (than occupied site densities), large bursts will necessarily become limited favoring smaller bursts. Since this happens at a smaller length scale for larger correlations, we expect to see a preference for smaller bursts which is what we find since $\tau \rightarrow 2$ for $H \rightarrow 0$. 

In nature, complex systems may posses multiple correlation lengths dominating behavior within their respective regimes. 
We have shown that depending on the growth dynamics, long-range order can subsequently be modified through the competition of correlation mechanisms and alter the scale invariant behavior in non-trivial ways. Contrary to finite size scaling requirement, we find the existence of multiple correlation lengths to be consistent with power-law behavior. Namely, we observe the ability of a stochastic process to manifest itself across a wide range of scales, and in some cases destroying fractal scaling behavior, and in other cases preserving it. Thus, understood across all length scales, the correct approach is perhaps a multifractal one where the characteristic distributions behave with moment description, $M_k(L) \sim L^{y(k)}$, with the key additional understanding being that for many of the scales of interest scale invariance is essentially preserved.

The multifractal framework \cite{halsey1986fractal} understood through the lens of the heirarchy of correlation lengths would have a set of fractal scalings describing the dominant singular behavior associated with each length scale. Briefly, in the case of a single length scale, which in addition gives rise to hyperscaling, yields moment distributions in terms of correlation lengths according to, 
\begin{equation}
M_k(l, \xi) \sim \xi^{(1+k+\tau)D_f} f(l/\xi) 
\end{equation}
where $k$ represents the k-th moment,$D_f$ the characteristic mass scaling, and $f(l/\xi)\rightarrow 1$ for $l\ll \xi$. The essential behavior is that exponents of successive moments are equally spaced according $kD_f$, since $k$ is an integer. Should there exist multiple scaling regimes, $\xi_b, \xi_H$ such that the burst distribution behaves differently for respectlively length scales, then we expect the moment behavior to follow $M_k \sim \xi_b^{(1+k+\tau)D_f} (L/\xi_b)^{(1+k+\tau_H)D_H}$. In terms of the two length scales this becomes $M_k \sim \xi_b^{y_b(k)}\xi_H^{y_H(k)}$. In principle the behavior of $y_b(k),y_H(k)$ can be determined from the system and therefore of our model provides an excellent case study for developing and better understanding current multifractal analysis techniques. Since multifractality is generally understood to arise in the presence of a novel or unknown relation between a spectrum of scaling exponents and successive moments, the approach informed by a correlation length analysis would be to identify the relevant correlation lengths in the system and establish whether multifractality necessarily emerges from a multiplicity of correlation lengths.
\section {Discussion}
In our previous study \cite{ortez2023avalanche}, we found that the AIP model was a pseudo-critical model possessing multiple correlation lengths. 
In this study, we include the addition of long-range order emerging from site strength Hurst correlations. This modification significantly alters the clustering behavior of sites into bursts, consequently affecting the critical Fisher distribution. We observe a clearer distinction in scaling behavior across different scaling regimes. Specifically, for length scales up to $l<\xi_b$, we observe different classes of meta-stable spinodal type growth governed by Hurst parameter $H$. More importantly, we establish that the existence of critical Fisher distribution $n_s(\tau, \sigma)$ remains, even when the underlying burst mechanism is not intrinsically critical (i.e., is not dictated by the critical parameter, $\epsilon_T$). In this regime, the Hurst long-range order is found to maintain scale-invariant bursts across a wide spectrum of length scales, irrespective of critical control parameter $\epsilon_T$. 

With these additions, we have been able to distance the behavior of correlated AIP from traditional critical mechanisms. This offers some interesting possibilities. First, correlated AIP rather uniquely address concerns with both SOC and traditional critical approaches. One of the criticisms of invoking critical behavior is the requirement of a finely tuned balance between small-scale order and large-scale disorder by mandating the system be near the critical point. This raises the question of whether traditional critical theory can fully encompass all forms of emergent long-range order--a scenario that could be infrequent, given its reliance on a singular point in phase space. Conversely, SOC systems are said to have offered the advantage that they do not require systems to be finely tuned to a particular value of phase space. Rather, a small external driving mechanism allowed the system to form meta-stable growth dynamics which results in the system innately growing in the critical regime. However, authors have questioned whether SOC systems need not be fine tuned \cite{vespignani1997order}. For example, \cite{grassberger1996self} shows how in SOC singularities arise not from order parameters, but instead, from control parameters which have a critical value. Also, \cite{gabrielli2000invasion} argues for the existence of characteristic ratio driving the behavior of SOC systems. 
 
While our findings support these claims (in random AIP, the system needs to be near the critical threshold/driving ratio), here, we also show how scale invariant behavior resulting from the competition of correlation mechanisms uniquely alleviates many of these concerns. Ever since Fisher showed that much of critical behavior could be characterized by critical distribution, $n_s(\tau, \sigma)$ \cite{fisher1967theory}, it has been widely assumed that the existence of Fisher type of distribution demonstrated critical behavior. In this study, we find this to strictly not be the case. We find a Fisher distribution even when the driving mechanism is not near its critical value. That is, the Fisher distribution exists largely independent of its proximity to the critical value (in our case the critical value is the burst threshold that becomes increasingly degenerate as $H\rightarrow0$). Therefore, our correlated AIP model avoids the necessity of applying artificial phase transition mechanics or arguing for self-organization around a general critical point. Thus, in systems with implicit long-range correlations, the emergent long range order of small scale stochastic dynamics can be strongly influenced by implicit long range correlations, and in such a way that preserves dynamic scale invariant properties (ie burst/cluster formation) without requiring any fine tuning or control parameters.

The generality of this result applies directly to broad extrema SOC type systems(of which AIP and CAIP belong), and can naturally be applied to stochastic energy minimization systems like interface motion in disordered media leading to domain walls \cite{cieplak1994optimal}, minimum spanning trees describing strongly disordered spin-glass models \cite{jackson2010theory}, abrupt species morphology changes through gradual changes in biologic fitness \cite{sneppen1995evolution}, and optimal neural topologies \cite{bornholdt2000topological}.

However, as much of our work is focused on the features of the IP process, which is best known as a drainage process of fluid infiltration \cite{stark1991invasion,sornette2006critical,knackstedt2021invasion,klein2007structure}, we focus on CAIP's application to fracture mechanics and induced seismicity\cite{norris2014loopless,rundle2020constrained,ortez2021universality,ortez2023avalanche}.

In \cite{chen1991self}, the authors argued for SOC description of tectonic seismicity producing rupture events with $b\sim 0.4$. This follows the work of \cite{vere1976branching, bebbington1990percolation} among others, that the comparison of model event scalings should be made independent of the $3/2$ energy scaling factor implicit in G-R scaling values. Without this factor, tectonic seimicity is described by $b \sim 2/3$, which is closer to many mean-field models $b=1/2$ and our random AIP model,  $b\sim0.6$. However, with correlations, we can obtain $b$-values in the range $[0.6,1.0]$ depending on the $H$. This can in part account for the larger $b$-values associated with induced seismicity, $[0.8,1.3]$. 

However, given the wide range of observed induced seimic event scalings, it is likely necessary to account for the inherently 3-d injection activity that only in some cases can be constrained to be 2-d. \cite{maxwell2011microseismic} shows how even within the same shale, the "effective" dimension of the injection activity can greatly differ. A previous study \cite{norris2015anisotropy} attempted to parameterize 1d and 2d growth through an anisotropic preference for growth along one of the axes. In the limit where growth was strongly directed and along 1 axes, the burst scaling changed from $\tau(2D)=1.483 \rightarrow \tau(1D)=1.451$. This represents a change far too small to account for the diversity of $b$-values associated with induced seismicity if only the dimension is allowed to change. Merely increasing the dimension of our random AIP model is unlikely to have the requisite impact on the burst scaling. This likely occurs because, with the introduction of a new degree of freedom, all burst sizes have an almost equal probability of increasing in size for scale invariant systems. Also, for scalings larger than mean-field, a higher dimension generally results in a lower 'b-value', which means larger clusters are favored. Beyond the critical dimension (d=6), we anticipate that all scaling will conform to mean-field theory suggesting that as the dimensionality increase $\tau \rightarrow 3/2$, and whose scaling is certainly inconsistent with observed induced seismicity. 

This problem is potentially side-stepped if we extend Hurst site correlations to 3d as well. Since burst epicenters are primarily distributed according to typical percolation process on small length scales, However, they exhibit crossover behavior for length scales greater than $\xi_b$, where the distribution of burst centers resembles that of a random walk. As a result, clusters larger than $\xi_b$ do not adhere to the scaling typical of percolation. Instead, they tend towards a uniform density. This inherent limitation naturally reduces the probability of forming larger clusters

In 2d we have already found that larger Hurst correlations prefer growth by smaller bursts, and when we allow this preference to be amplified by an additional degree of freedom, its likely that we will observe an even more dramatic change in burst scaling. We know that the random walk nature of bursting behavior is essentially preserved with the introduction of $H$, and permits a higher density of bursts then would be allowed if it were entirely a percolation system. This can account for burst densities which are essentially uniform on large scales, and have mass scaling that is given by the usual non-fractal dimension scaling, $d$. Thus, we propose that a future study extending both the AIP model and Hurst correlations into 3d should better be able to account for the range and magnitude of $b$-values associated with induced seismicity. 

\section{Acknowledgements}
The research of RAO and JBR has been supported by a grant from the US Department of Energy to the University of California, Davis. DOE Grant No. DE-SC0017324.

\bibliography{CAIP_bibliography}
\appendix

\section{LRC and extended Harris Criteria}
\label{appedix:LRC}
Naively, one might fail to properly appreciate the unique impact of long-range correlations on critical behavior, since one might well consider any other kind of change to the site lattice structure and consider its effects. However, as Harris \cite{harris1974effect} found in considering the effects of random defects on the critical temperature of the Ising model, the only defects that can have an effect are those whose correlation length, $\xi_H$, is comparable to the correlation length of the unmodified lattice, $\xi$. Thus, since near the critical point $\xi$ is described by a power-law, only those defects whose statistics similarly produce long-range correlations could have any effect on the critical behavior. Any short-range correlations would fail to meet this criterion. This reiterates the focal feature of critical behavior, where small-scale interactions can eventually become renormalized, and only those that persist on all scales contribute to its behavior.

We consider correlations that are sufficiently long-ranged while also convergent for all distances, and whose auto-correlation function is given by,
\begin{equation}
C(r) \sim r^{-a}
\label{eq:LRCfuntion}
\end{equation} 
where $r$ is the distance between sites, and $a$ is less than dimension $d$. Since this is the percolation problem, the auto-correlation function describes the correlations in site occupation, that is, the likelihood that two sites a distance $r$ are occupied. The long-range correlations are therefore an additional mechanism contributing to the site occupation probability other than the usual uniform occupation probability, $p$. 

We can calculate how these kinds of site strength correlations affect the fluctuations in the control parameter, $\langle \delta T^2 \rangle$,  according to 
\begin{equation}
\begin{split}
\langle \delta T^2 \rangle &\sim \xi^{-d} \int_0^\xi dr \ C(r)r^{d-1} \\
&= \xi^{-d} \int_0^\xi dr \ r^{-a+d-1} \\
&\sim \xi^{-a}
\end{split}
\end{equation}
where again, we can define a correlation length characterizing the average spatial extent of fluctations given by, $\xi^{-a}$.

If the system is still to have a single uniform critical transition, then it should be the case that these fluctuations produce a correlation length less than that of unmodified transition. That is the fluctuations should be less than critical fluctuations, leading to the condition,
\begin{equation}
\begin{split}
\frac{\langle \delta T^2 \rangle}{(T_c-T)^2} &\sim (T_c-T)^{a\nu-2} \\
&\rightarrow 0
\end{split}
\end{equation}
where we made use of the relation, $\xi \sim (T_c-T)^{-\nu}$ to expand the ratio. For the ratio to go to zero near the critical point we require the exponent to be greater than zero. This leads to the condition on $a$ for the largest value of long-range correlations such that it will affect the critical transition while preserving the existence of a uniform transition. This is given by,
\begin{equation}
a\nu - 2 > 0
\label{eq:HarrisCond}
\end{equation} 
Thus, we can expect changes to the critical behavior if $a<2/\nu$. For our AIP model $\nu \approx 1.30$, we should expect that the minimum value requires $a<3/2$.

\section{Fourier Filter Correlation Method} 
\label{appendix:FF Correlation Method}
It is common to parameterize such long range scale invariant correlations using the Hurst exponent, where the (auto)correlation function, $C(r)$ defined as $C(r)=\left\langle u(r^\prime)u(r+r^\prime)\right\rangle$ has the following behavior:
\begin{equation}
C(r) \propto r^{2\alpha}
\label{eq:hurstExp}
\end{equation}
The Hurst exponent is given by $H=2\alpha$ and allowed to take on values in range [0,1]. Behavior of the correlations are antipersistent for $H<1/2$ and persistent for $H>1/2$. For $H=1/2$, the statistics follow fractional Gaussian noise, being neither persistent nor antipersistent.

There are a number of techniques for simulating fractional Brownian statistics \cite{fisher2012science}. We use the Fast Fourier transform(FFT) filter technique because of its computational efficiency. This technique relies on imprinting the desired correlations in the Fourier wave vector space, $\vec{k}$, and then applying an inverse FFT(IFFT) to create a lattice with correlated sites of form Equation \eqref{eq:hurstExp}. Formally, we will be working with 2 dimensional Fourier transforms, and it is well known that the Fourier transform of the autocorrelation function gives the Fourier power spectral density. That is, the correlation function, $\langle u(\vec{x})u(\vec{x}+r)\rangle$ and the power spectral density $S(\vec{k})$ are related according to:
\begin{equation}
\langle u(\vec{x})u(\vec{x}+r) \rangle=\int_{\mathcal{R}^n}S(\vec{k})e^{-i2\pi  \vec{k} \cdot \vec{x}}d\vec{k}
\label{eq:corrFcnFT}
\end{equation}
We can make use that we are only concerned with the distance between two points. This leads to a suitable definition of a radial wave vector defined as $k_r = \sqrt{1+s^2+t^2}$ and with a switch of coordinates allows us to write it as a one dimensional Fourier Transform. 
\begin{equation}
C(r)=\int S(k_r)e^{-i2\pi k_r r}2\pi k_r dk_r
\end{equation}
To create correlations of the form Equation \eqref{eq:hurstExp}, our power spectral density should be made to follow the following power-law:
\begin{equation}
S(k_r) \propto \frac{1}{k_r^\beta}
\label{eq:pwrSpectrum}
\end{equation}
To relate the exponents between Equation\eqref{eq:hurstExp} and Equation\eqref{eq:pwrSpectrum} we can solve Equation\eqref{eq:corrFcnFT} after substituting Equation\eqref{eq:pwrSpectrum} which gives the following integral to be solved:
\begin{equation}
C(r) =2\pi \int k_r^{-\beta+1}e^{-2\pi i k_rr}dk_r
\label{eq:corrIntegral}
\end{equation}

To solve the above integral we first make use of the following relation:
\begin{equation}
\frac{1}{k^\beta} =\frac{2\pi^{\beta/2}}{\Gamma(\beta/2)}\int_{0}^{\infty} \lambda^{\beta -1}e^{-\pi \lambda^2k^2} d\lambda
\label{eq:pwrLawExp}
\end{equation}
The right side is easily Fourier transformed and upon switching the order of integration, we get:

\begin{equation}
\int_{\mathcal{R}}e^{-\pi \lambda^2|k|^2}e^{-2\pi i kr} dk = \lambda^{-1}e^{-\pi {|r|}^2/\lambda^2}
\label{eq:gaussianFT}
\end{equation}
Then taking the 2D Fourier transform of both sides and plugging into Equation\eqref{eq:pwrLawExp} gives:
\begin{equation}
\begin{split}
\int_{\mathcal{R}} k^{-\beta+1}e^{-i2\pi  k_rr} dk_r &= \frac{2\pi^{\beta-1/2}}{\Gamma(\beta-1/2)} \int_{0}^{\infty}d\lambda {\lambda}^{\beta-2}\left[\lambda e^{-\pi{|r|}^2/\lambda^2}\right] \\
&=\frac{2\pi^{\beta-1/2}}{\Gamma(\beta-1/2)} \int_{0}^{\infty}d\lambda {\lambda}^{(\beta-2)-1}e^{-\pi{|r|}^2/\lambda^2} \\
&=\frac{2\pi^{\beta-1/2}}{\Gamma(\beta-1/2)} \frac{\Gamma((\beta-2)/2)} {2\pi^{1/2-\beta+1/2}} \frac{1}{{|r|}^{1-\beta+1}} \\
&\propto r^{\beta - 2} \\
\label{eq:intSolve}
\end{split}
\end{equation}
Setting the exponents equal between the final line of Equation \eqref{eq:intSolve} and Equation \eqref{eq:hurstExp} gives:
\begin{equation}
2\alpha=\beta-2
\label{eq:genHurst}
\end{equation}
This gives our final relationship between the Hurst exponent and the appropriate Fourier power spectrum filter function exponent.
\begin{equation}
\beta = 2(\alpha+1)
\label{eq:betaHurstRelation}
\end{equation}

\section{Correlation Algorithm}
We execute this Fourier filter technique using an FFT on a NxN dimensional array with complex coefficients \cite{turcotte1997fractals}. The algorithm is outlined as follows:
\begin{enumerate}
	\item We generate a NxN array with each value,$h_{nm}$, assigned a random value from a Guassian probability distribution.
	\item We execute a 2D Fast Fourier Transform(FFT) giving an array of complex coefficients, $H_{st}$.
	\item We define radial wave number $k_r$, which is non-zero for $s=t=0$, as follows:
	\begin{equation}
	k_r=\sqrt{1+s^2+t^2}
	\end{equation}
	\item Since $S(k_{st}) \propto {|H_{st}|}^2$ we define a new set of complex coefficients, $H_{st}$, multiplied by the appropriate filter function:
	\begin{equation}
	H_{st}^\prime = H_{st}/k_r^{\beta/2}
	\end{equation}
	\item Apply an inverse FFT(IFFT) on $H_{st}^\prime $ to produce a new NxN array with coefficients, $h_{st}^\prime$ with the desired correlations.
	\item Apply the error function, $erf(h_{st}^\prime)$, to return a uniform correlated distribution with values in range [0,1].
\end{enumerate}

We illustrate an example of the types of correlations produced by our algorithm in Figure \ref{fig:corrLattEx}.
\end{document}